\documentclass[12pt]{article}%
\usepackage{amssymb}
\usepackage{amsmath}
\usepackage{amsthm}
\usepackage{amsfonts}
\usepackage{graphicx}
\usepackage{subfigure}%
\input epsf.tex
\begin{document}
\bigskip\begin{titlepage}
\begin{flushright}
UUITP-09/05\\
hep-th/0506219
\end{flushright}
\vspace{1cm}
\begin{center}
{\Large\bf 4D black holes and holomorphic factorization of the 0A matrix model\\}
\end{center}
\vspace{3mm}
\begin{center}
{\large
Ulf   H.\   Danielsson{$^1$}, Niklas Johansson{$^2$}, Magdalena Larfors{$^3$},  \\ Martin  E.
Olsson{$^4$} and Marcel Vonk{$^5$}} \\
\vspace{5mm}
Institutionen f\"or Teoretisk Fysik, Box 803, SE-751 08
Uppsala, Sweden \\
\vspace{5mm}
{\tt
{$^1$}ulf.danielsson@teorfys.uu.se\\
{$^2$}niklas.johansson@teorfys.uu.se\\
{$^3$}magdalena.larfors@teorfys.uu.se\\
{$^4$}martin.olsson@teorfys.uu.se\\
{$^5$}marcel.vonk@teorfys.uu.se\\
}
\end{center}
\vspace{5mm}
\begin{center}
{\large \bf Abstract}
\end{center}
\noindent
In this letter, we relate the free energy of the 0A matrix model to
the sum of topological and anti-topological string amplitudes. For 
arbitrary integer multiples of the matrix model self-dual radius we
describe the geometry on which the corresponding  topological string
propagates. This geometry is not the one that follows from the usual 
ground ring analysis, but in a sense its ``holomorphic square root''. Mixing of 
terms for different genus in the matrix model
free energy yields one-loop terms compatible with type II strings on
compact Calabi--Yau target spaces. As an application, we give an
explicit example of how to relate the 0A matrix model free energy to that of a
four-dimensional black hole in type IIB theory, compactified on a
compact Calabi--Yau. Variables, Legendre transforms, and large
classical terms on both sides match perfectly.  \vfill
\begin{flushleft}
June 2005
\end{flushleft}
\end{titlepage}\newpage

\section{Introduction}

Recently, a very interesting relation between four-dimensional ${\cal N} = 2$ supersymmetric BPS black holes and
topological strings has been proposed \cite{Ooguri:2004zv}. This
correspondence relates the black hole free energy in type IIB
string theory, compactified on a Calabi--Yau threefold, to the topological and anti-topological string amplitudes on this same manifold,
according to
\begin{equation}
\mathcal{F}_{BH}=\mathcal{F}_{top}+\overline{\mathcal{F}}_{top}.
\end{equation}
In this relation, the complex structure moduli of the Calabi--Yau are fixed in terms of the black hole charges by certain attractor equations.

Topological strings are often related to matrix models. It is well known that the topological theory on the conifold is
perturbatively equivalent to $c=1$ bosonic non-critical string theory at self-dual
radius, and hence to a matrix model \cite{Ghoshal:1995wm}. In fact, there are
quite general correspondences between matrix models and topological strings on
non-compact Calabi--Yaus based on ground ring considerations
\cite{Witten:1991zd,Aganagic:2003qj, Ita:2004yn, Hyun:2005fq}.

Usually, the advocated correspondence relates the matrix model free energy
$\mathcal{F}_{MM}$ directly to the topological string amplitude $\mathcal{F}%
_{top}$. However, as is by now well known, the exponential of $\mathcal{F}_{top}$ should not be viewed as a partition function, but rather as a wave function \cite{Witten:1993ed,Dijkgraaf:2002ac,Vonk:2003th}. Thus, it seems unnatural to relate
$\mathcal{F}_{MM}$ directly to $\mathcal{F}_{top}$. If $\mathcal{F}_{MM}$ is a
true free energy one should rather make the identification\footnote{Note that
this does not mean that the identification of the $c=1$ free energy with the
topological string amplitude is wrong, since in that case one divides
the natural (and real) free energy by two so to make it agree with 2d
space-time calculations. However, the result is only true for a particular choice of polarization of the phase space $H^3(CY)$.}
\begin{equation}
\mathcal{F}_{MM} = \mathcal{F}_{top} + \overline{\mathcal{F}}_{top},
\end{equation}
thereby directly relating the free energy of the matrix model to the one for
the black hole. We will see many indications in this paper that this is the
right way to think of the relation between matrix models and topological
strings. In particular, as we will show in an explicit example in section
\ref{sec:bhentropy}, the black hole free energy resulting from a deformed conifold with complex deformation parameter is precisely given by the free energy of the 0A matrix model at
the self-dual radius. From the construction we present in section
\ref{sec:geomdef}, the generalization to $n$ times the self-dual radius, and
its interpretation in terms of $n$ conifolds, is straightforward\footnote{This
will also serve to clarify some of the results in \cite{Danielsson:2004ti},
where the issue of associating specific radii to a certain number of conifolds
was discussed.}. The underlying property of the 0A model which makes this
interpretation natural is its holomorphic factorization, as we will discuss in
section \ref{sec:geomdef}.

Note that the identification of the 0A model at self-dual radius with a {\em single} conifold is different from what one naively may expect from ground ring relations. The reason for this is exactly the holomorphic factorization, which forces us to look at a manifold which in a sense is also the ``holomorphic square root'' of the one given by the defining relation of the matrix model ground ring. By studying the 0A and 0B
matrix models \cite{Takayanagi:2003sm, Douglas:2003up} at various radii, and
investigating their relation to the $c=1$ matrix model, we show how these new geometries
are constructed. 

The procedure also gives new insights on the genus expansions
of the matrix model free energy. In particular there will be a crucial mixing between genus 0 and genus 1
terms. The result of the mixing is that the genus 1 term in the expansion of the free energy will have the numerical coefficient $-1/12$ for both the holomorphic and the anti-holomorphic term. As explained in
\cite{Vafa:1995ta}, and later pointed out in \cite{Ghoshal:1995wm}, this is
what is required for string theory to resolve the singularity associated to the shrinking of
a cycle in the geometry. Again, this result is different from what one may expect from a naive ground ring analysis.

We then turn to the correspondence between matrix models and black holes. As
has been argued recently \cite{Danielsson:2004ti}, the relations mentioned
above allow for a description of the black hole entropy\footnote{There has been some discussion recently on the question whether the quantity calculated in \cite{Ooguri:2004zv} should really be called an entropy, or rather an index. We will use the term ``entropy'' throughout this paper, but the reader should be aware that this term is not to be taken too literally.} in terms of the free
energy of a matrix model. However, there are some important issues that need
to be clarified in the proposed correspondence. Most importantly, in
\cite{Danielsson:2004ti} non-compact Calabi--Yaus are considered as internal
spaces for string compactification. When we view these as local models for compact Calabi--Yaus, we would like to think of all parameters of the noncompact Calabi--Yau as moduli. Hence, the manifold must have at least one A-cycle at infinity, and it is not
straightforward to obtain the dependence of the black hole entropy on all the
charges. In the present letter we remedy this deficiency by
treating compact Calabi--Yaus. It may sound strange that a matrix model can say something about string theory on a compact Calabi--Yau. The reason this happens here is that one can find special charge configurations which, through the attractor equations of \cite{Ferrara:1995ih,
Strominger:1996kf}, result in a singular compactification space only at the horizon. This is enough to allow one to calculate the black hole free energy from the matrix model.

Thus, our internal space is a truly compact manifold which only near the black hole horizon ``decompactifies'' into a conifold-like geometry. Note that this decompactification only involves quantities that are expressed in terms of the complex structure moduli. In particular, it seems perfectly possible to keep the K\"ahler volume of the Calabi-Yau finite throughout space-time. Since it is this volume which appears in the four-dimensional Newton's constant, we can really speak of four-dimensional gravity with nonzero coupling constant (and hence, for instance, of true black holes) in this context.

The paper is organized as follows. In section \ref{sec:geomdef} we introduce
the matrix models and describe the new geometrical interpretation of the
0A matrix model. We explain the crucial mixing of genus 0 and genus
1 terms that takes place at multiples of the self-dual radius, and its
implications for which geometry one should consider. We also work out the case
of fractional radii. In section \ref{sec:bhentropy} we discuss the
correspondence between black holes and matrix models. We describe how the
attractor equations can ``decompactify'' a compact internal space on the black
hole horizon. As an explicit example, we study a Calabi--Yau with a conifold
point. We match variables, Legendre transforms, and large classical terms on
the matrix model and black hole sides. Finally, we summarize and discuss our
results.

While this manuscript was being prepared for submission, we received the interesting paper \cite{Maldacena:2005he}, which discusses the 0A and 0B matrix models in a lot of detail. Also in that paper, the holomorphic factorization of the matrix model partition functions plays an important role.

\section{The geometry of the 0A matrix model}
\label{sec:geomdef}
This section describes the geometrical interpretation we propose for the
0A matrix model. Let us begin by explaining the matrix model nomenclature we use. 
In the eigenvalue description, the $c=1$ matrix model describes free fermions in an inverted
harmonic oscillator potential, with the Fermi sea filled on one side of the potential. This
model is nonperturbatively unstable due to tunnelling. The 0A matrix model\footnote{The particular model we study was introduced and further studied in \cite{Jevicki:1993zg}-\cite{Danielsson:2004xf}, and is also known in the literature as the ``deformed matrix model''. It can be shown by integrating out eigenvalue phases that the gauged and {\em holomorphic} matrix model that is often used to describe the two-dimensional type 0A string theory is equivalent to this Hermitean matrix model.} differs from this model by a term $M/x^{2}$ which is added to the potential.
This deformation effectively removes one side of the potential, thus creating a stable model with one Fermi
sea. One could also consider the undeformed matrix
model with both sides of the potential filled. This corresponds to the 0B matrix model, which also is non-perturbatively stable. The 0A and 0B matrix models were constructed in \cite{Takayanagi:2003sm,Douglas:2003up}, and their relations at different
radii were discussed in detail in \cite{Danielsson:2004ti}, to which we refer
for further reading. 

We will mainly be interested in matrix model free energies. Unless stated
otherwise, the free energies are given in the grand canonical ensemble. We use
the notation
\begin{equation}
\mathcal{F}_{MM}=-2\pi\beta RF_{MM}=\ln\mathcal{Z}_{MM},
\end{equation}
where $MM$ can stand for ``$c=1$'', ``0A'' or ``0B''. ${\mathcal Z}_{MM}$ and $F_{MM}$ are the usual partition function and free energy of the matrix model. In the case of the $c=1$ matrix model, we have
\begin{equation}
\mathcal{F}_{c=1}(\mu,R)=\operatorname{Re}f(i\mu,R)
\end{equation}
with%
\begin{equation}
f(i\mu,R)  =\sum_{n,m=0}\ln\left(  \frac{2n+1}{2}+\frac
{2m+1}{2R}+i\mu\right)  .
\end{equation}
The genus expansion becomes%
\begin{equation}
\mathcal{F}_{c=1}(\mu,R)=-\frac{R}{2}\mu^{2}\ln(\mu)-\frac{1}{24}\left(
R+\frac{1}{R}\right)  \ln(\mu)+... \label{eq:bosonicexp}%
\end{equation}
Throughout the paper, we use units where $\alpha^{\prime}=1/2$ on the 0A and
0B sides and $\alpha^{\prime}=1$ on the $c=1$ side. This means that the
various self-dual radii are $R_{SD}^{A}=\frac{1}{2},R_{SD}^{B}=1$ and
$R_{SD}^{c=1}=1$, respectively.

\noindent
The expansion for the 0A free energy \cite{Danielsson:2004ti} is given by
\begin{eqnarray}
\mathcal{F}_{0A}(\mu,R,q) &  = & 2\mbox{Re}\Big[f\Big(\frac{q+i\mu}{2},2R\Big)\Big] \nonumber \\
&  = & 2\mbox{Re}\Big[  R(q/2+i\mu/2)^{2}\ln(q/2+i\mu/2) \nonumber  \\ 
&& -\frac{1}{24}\Big(
2R+\frac{1}{2R}\Big)  \ln(q/2+i\mu/2)+\ldots\Big]
\label{FA=Fbos}%
\end{eqnarray}
where $q$ is related to the coefficient of the deformation term in the
potential as $q^{2}\nolinebreak = \nolinebreak M+1/4$. In the corresponding two-dimensional string theory, $q$ is the net amount of D0-brane charge in the background. The above formula explicitly displays the
holomorphic factorization mentioned in the introduction. In a very precise
sense, the 0A partition function is the holomorphic square of the
\textquotedblleft complexified\textquotedblright\ $c=1$ partition function.

Let us explain some of the perhaps strange-looking factors of $i$ in the above formulae. In the literature one usually encounters expansions in the parameters $\mu$ or $\mu+iq$. However, as emphasized in
\cite{Danielsson:2004ti}, it is really the sign in front of $\mu$ that changes
when taking the complex conjugate. This may seem like an academic point since
all the signs are going to be squared away anyway. However, it will be
important when we construct the geometries, and natural later on when we are
matching variables. Moreover, it \emph{will} matter for the nonperturbative
part of the theory \cite{Danielsson:2004ti}.

\subsection{The 0A matrix model at self-dual radius}

By going to a double scaling limit, it has been shown that the free energy
of the $c=1$ matrix model at self-dual radius is identical to the
topological string amplitude on the conifold \cite{Ghoshal:1995wm}. The two
terms in Eq.\ (\ref{eq:bosonicexp}) then correspond to the genus 0 and genus 1 terms
of the topological string. From Eq.\ (\ref{FA=Fbos}) we also see that at
self-dual radius, the free energy of the 0A matrix model is identical to
the sum of the topological and anti-topological amplitudes of the
conifold. We have the relation
\begin{equation}
\mathcal{F}_{0A}(\mu,R_{SD}^{A},q)=2\mbox{Re}\left[  f\left(\frac{q+i\mu}{2}%
,R_{SD}^{c=1}\right)\right]  =2\mbox{Re}\left[  {\mathcal{F}}_{top}\left(\frac{q+i\mu}%
{2}\right)\right]  .\label{FA=Ftop}%
\end{equation}
Thus, we relate the 0A
matrix model at self-dual radius to the conifold, with the equation
\begin{equation}
uv+(\mu - iq)=st.\label{conifold}%
\end{equation}
Note that the 0A theory has enough real parameters to describe one complex modulus. Only in the above expression and similar ones that follow, in order to make contact with existing literature, we use the ``conventional'' notation where $\mu$ is the real part of the parameter and $q$ the imaginary part. Note that we can do this without loss of generality, since we can always for example rescale $u$ and $s$ by a factor of $i$. 

Of course, since we take the real part of $f$, we could just as well have written it as a function of $q - i \mu$, leading to a conifold of the form
\begin{equation}
 uv + (\mu + iq) = st.
\end{equation}
Now, we can see how these two manifolds are related to the usual ground ring geometry. It has been proposed \cite{Ita:2004yn,Danielsson:2004ti} that the ground ring equation for the 0A model at self-dual radius including both its deformations is
\begin{equation}
 (uv+ \mu)^2 = st - q^2.
\end{equation}
We can rewrite this as
\begin{equation}
 (uv + \mu - iq)(uv + \mu + iq) = st.
 \label{eq:doublecon}
\end{equation}
It is useful to view this geometry as a fibration over the $uv$-plane. The fiber $st =$ \nolinebreak const is a cylinder, except over the loci $uv + (\mu \pm iq) = 0$, where it is the intersection of two complex planes in a single point. As is well-known (see the appendix of \cite{Danielsson:2004ti} for a pictorial explanation) the complex structure moduli of the manifold are related to A-cycles which are localized near these loci, and B-cycles which start there and run off to infinity. Since for $q$ large these loci are far away from each other, the geometry of Eq.\ (\ref{eq:doublecon}) then effectively reduces to two independent copies of the deformed conifolds we mentioned above. This is the intuitive reason why in the right polarization and perturbatively, the topological amplitude on the ground ring geometry and the sum of topological and anti-topological conifold amplitudes give the same result.

Since the 0A model is well defined beyond its perturbative expansion \cite{Takayanagi:2003sm,Douglas:2003up}, it would clearly be interesting to
further explore also its nonperturbative aspects at special radii.

\subsection{The 0A matrix model at other radii}

It is natural to ask whether the 0A matrix model at other radii
corresponds to topological string theories on other singular geometries. It was argued in \cite{Strominger:1995cz, Vafa:1995ta} that the type II string compactified on a compact Calabi--Yau with shrinking cycles is only non-singular if the corresponding topological string amplitude has a one-loop term coefficient of $-k/12,\;k\in\mathbb{Z}$. Eq.\ (\ref{FA=Fbos}) shows that, apart from at self-dual radius, the 0A matrix model does not satisfy this requirement. Hence, it seems that we are in big trouble if we want to identify the general matrix models we consider with topological strings --- in particular if we would like these topological strings to live on double scaling limits of compact manifolds, as we will in the next section. 

However, because of the double scaling limit the parameters in the model will be of the same order of magnitude as Planck's constant, and it is not immediately obvious anymore that one can match the expressions
on the matrix model and the topological string side genus by genus. In particular, it might be the case that the genus
0 term on the matrix model side contributes to the genus 1 term on the
topological string side. Similar types of genus mixing have previously been
considered in \cite{Danielsson:1993dh, Kapustin:2003hi}. Below we argue that
this is the correct way of viewing the matrix model free energy.

As a motivation we show that, in order for the 0A expression to reduce to 0B
as $M\rightarrow0$, we need to make such a reinterpretation of terms.
If we plug in $q = \frac{1}{2}$ in Eq.\ (\ref{FA=Fbos}) we get an expression
\begin{equation}
\label{FA_q=1/2}%
\begin{split}
\mathcal{F}_{0A}(\mu,{R},\frac{1}{2}) = 2\mbox{Re}\left[  \frac{R}{4}(\frac{1}{4} - \mu
^{2})\ln(\mu+i\frac{1}{2}) - \frac{1}{24}(2R+\frac{1}{2R}%
)\ln(\mu+i\frac{1}{2}) +...\right]  ,
\end{split}
\end{equation}
where we have skipped imaginary and analytic terms. We see that we cannot
match the genus 0 (1) term in this expression to the genus 0 (1) term of the
0B free energy directly. For example, the self-dual radius of the 0B model is 1,
while in this expression it appears to be $1/2$. However, if we move the
$\frac{R}{16}\ln(\mu+iq)$ from the genus 0 term to the genus 1 term in
Eq.\ (\ref{FA_q=1/2}), we indeed get the expression for half the 0B free
energy\footnote{We only get half since in the 0A theory, half of the states have to be removed \cite{Danielsson:2004ti}.}. 
Note that we also get the correct self-dual radius for 0B by making
this shift.

We now turn to the reinterpretation of terms suitable for describing
topological strings on Calabi--Yau manifolds. To this end we use the formula
of Gopakumar and Vafa \cite{Gopakumar:1998vy} for $\mathcal{F}_{c=1}(R_{SD}^{c=1}/n)$:
\begin{equation}
\mathcal{F}_{c=1}\left(  \mu,\frac{R_{SD}^{c=1}}{n}\right)  =\sum_{k=-(n-1)/2}%
^{(n-1)/2} \mathcal{F}_{c=1}\left(  \frac{\mu-ik}{n},R_{SD}^{c=1}\right)  .\label{bos_mR}%
\end{equation}
Using Eq.\ (\ref{FA=Fbos}) and going to $n$ times the self-dual radius, this
can be recast into a formula for $\mathcal{F}_{0A}$:
\begin{equation}
\mathcal{F}_{0A}(\mu,nR_{SD}^{A},q)=2\mbox{Re}\left[  \sum_{k=-(n-1)/2}%
^{(n-1)/2}f\left(  \frac{(q+\frac{2k}{n}+i\mu)}{2},R_{SD}^{c=1}\right)
\right]  .\label{multR}%
\end{equation}
The 0A free energy at $n$ times the self-dual radius is thus given as two
times the real part of the sum of $n$ $c=1$ free energies at self-dual
radius. Since the coefficient in front of the 1-loop term of each
$f$ is $-1/12$, this immediately shows that we have succeeded
in rearranging the terms so that they make sense from a type II string theory point of
view. It also means, by the result of Ghoshal and Vafa \cite{Ghoshal:1995wm},
that it computes a sum of $2\mbox{Re}\mathcal{F}_{top}$ on $n$ conifolds.

We are now in a position to say something about the geometrical interpretation of
the 0A matrix model at $n$ times the self-dual radius. It should
correspond to a certain double scaling limit of the topological theory on a
Calabi--Yau with $n$ three-cycles that can shrink at different loci in moduli
space. Call the distance to these loci $t^{k}$,\; $k=-\frac{n-1}{2}%
,\ldots,\frac{n-1}{2}$. Then the limit described by the 0A matrix model
is $t^{k} \rightarrow0$ and $g_{top} \rightarrow0$ with
\begin{equation}
\label{moduli}\frac{t^{k}}{g_{top}}= \left(  q+\frac{2k}{n} \right)  + i\mu
\end{equation}
kept fixed\footnote{Eq.\ (\ref{moduli}) is the correct equation if the
parameters $t^{k}$ are chosen so that all $(t^{k})^{2} \ln t^{k}$ terms in
$\mathcal{F}_{top}$ appear with the same coefficient, and up to an overall normalization.}. Thus the Calabi--Yau
must allow for all cycles to shrink simultaneously. Note also that the matrix
model, having only two parameters, describes a very special limit of this geometry.

There are of course many geometries satisfying these properties, but some have a more natural interpretation than others. As an example, let us work out the geometry in more detail for the case $n=2$. At twice
the self-dual radius, the free energy is (recall that $R_{SD}^{c=1}%
=2R_{SD}^{A}=1$)
\begin{align}
\mathcal{F}_{0A}(\mu,1,q)  & =2\mbox{Re}\left[  f\left(  \frac{q-\frac{1}%
{2}+i\mu}{2},1\right)  +f\left(  \frac{q+\frac{1}{2}+i\mu}{2},1\right)
\right] \nonumber \\
& =2\mbox{Re}\left[  \mathcal{F}_{top}\left(  \frac{q-\frac{1}{2}+i\mu}%
{2}\right)  +\mathcal{F}_{top}\left(  \frac{q+\frac{1}{2}+i\mu}{2}\right)
\right]. \label{doubleR}
\end{align}
We see that there are two loci in parameter space where the corresponding
Calabi--Yau should have conifold singularities. Note that, again, this is half the number one would expect by a naive ground ring analysis.
Following the arguments in \cite{Gopakumar:1998vy}, such a manifold can be created by modding out the conifold $st = uv + (\mu - iq)$ by a $\mathbb Z_2$, changing variables from $(s^2, t^2)$ to $(s,t)$, and deforming the resulting $A_2$-singularity by $\pm \frac{1}{2}$, leading to
\begin{equation}%
\begin{split}
st &  =(uv+\mu-i\left(  q-1/2\right)  )(uv+\mu-i\left(  q+1/2\right)  )\\
&  =(uv+\mu)^{2}-2iq(uv+\mu)-M.
\end{split}
\label{manifold1}%
\end{equation}
Just as at the end of the previous section, this procedure boils down to simply multiplying the equations for the loci of the single conifolds. 

However, perturbatively we can rewrite Eq.\ (\ref{doubleR}) in several other ways, such as
\begin{align}
\ \mathcal{F}_{0A}(\mu,1,q)  & =2\mbox{Re}\left[  \mathcal{F}_{top}\left(
\frac{q-\frac{1}{2}+i\mu}{2}\right)  +\mathcal{F}_{top}\left(  \frac
{q+\frac{1}{2}-i\mu}{2}\right)  \right]  \\
& =2\mbox{Re}\left[  \mathcal{F}_{top}\left(  \frac{q-\frac{1}{2}+i\mu}%
{2}\right)  +\mathcal{F}_{top}\left(  \frac{-\left(  q+\frac{1}{2}\right)
+i\mu}{2}\right)  \right]  +\mbox{non-pert.} \nonumber
\end{align}
where in the first expression we have used $\mbox{Re}[f(z)]=\mbox{Re}[f(\bar{z})]$.
The second expression can be easily verified by examining the genus expansion. The
detailed form of the non-perturbative contribution was given in
\cite{Danielsson:2004ti} for the case of $q=0$. We might then conclude that the resulting
manifold is given by%
\begin{align}
st  & =\left(  uv+\mu-i\left(  q-1/2\right)  \right)  \left(  uv+\mu+i\left(
q+1/2\right)  \right) \nonumber \\
& =\left(  uv+\mu\right)  ^{2}+M+i\left(  uv+\mu\right). \label{manifold2}
\end{align}
We claim that this latter manifold is the more natural one corresponding to the ground state of the Fermi sea. (By general ground ring arguments \cite{Witten:1991zd}, one can argue that extra terms proportional to $uv$ and $1$ should correspond to excitations of the Fermi sea.) The reason for this is that we can now make contact with the higher
genus analysis described in \cite{Danielsson:1994ac}, which is equivalent to
the Kodaira-Spencer description of the topological string
\cite{Bershadsky:1993cx}. See also \cite{Aganagic:2003qj,Hyun:2005fq}, where similar techniques are used. To do this we need to consider
\cite{Danielsson:2004ti,Danielsson:1994ac} a superpotential given by
\begin{equation}
W=\frac{M}{D^{2}}+\left(  \frac{\mu}{D}-X+\sum t_{2k}D^{2k-1}\right)  ^{2},
\end{equation}
where $\left[  D,X\right]  =-i$. For the free energy we need not consider the
perturbations and we can put all $t_{2k}=0$. Hence, commuting everything to
the right, and putting $X=0$, we find
\begin{equation}
W=\frac{M}{D^{2}}+\frac{\mu^{2}}{D^{2}}-i\frac{\mu}{D^{2}}.
\end{equation}
We see that this matches the structure of the expression in Eq.\
(\ref{manifold2}) for $uv=0$ and up to an irrelevant complex conjugation. For
further details on how to perform the higher genus calculations in this
framework, see \cite{Danielsson:1994ac}. It would of course be very interesting to see if the manifolds that are natural from the Kodaira-Spencer point of view also allow for a more natural embedding into truly compact Calabi--Yaus.

For completeness, let us work out the case for fractional radii $R^{A}= \nolinebreak R_{SD}^{A}/n$. In this case, the 0A free energy should be written
\cite{Gopakumar:1998vy}
\begin{align}
\mathcal{F}_{0A}\left(  \mu,\frac{R^{A}_{SD}}{n},q\right)  &=2\text{Re}\left[f%
\left( \frac{q + i\mu}{2},\frac{R^{c=1}_{SD}}{n}\right)\right] \nonumber \\ &=2\text{Re}\left[  \sum_{k=-(n-1)/2}%
^{(n-1)/2}f\left(  \frac{q+2k+i\mu}{2n},R^{c=1}_{SD}\right)  \right]  .
\end{align}
In terms of the Calabi--Yau, a natural interpretation of this sum, which
contains $n$ terms, is that the total charge $q$ and potential $\mu$ is
associated to the $n$ conifolds in a specific way\footnote{Running slightly ahead of the black hole part of our story, let us make the following interesting observation. For the
$l^{\prime}$th conifold, say, the attractor equations \cite{Ferrara:1995ih,
Strominger:1996kf}, fix the complex structure moduli (including the imaginary part
\cite{Ooguri:2004zv}) at the horizon to
\begin{equation}
CX^{l+1}=\frac{1}{n}(2l+1+q+i\mu)-1,
\end{equation}
where $0\leq l\leq n-1$. It is interesting to compare this with the energy
eigenvalues of the 0A matrix model \cite{Danielsson:1993wq}:
\begin{equation}
E^{l}=i(2l+1+q + i \mu).\label{eq:energyeig}%
\end{equation}
Given our identifications, and the fact that the topological partition
function is peaked at the attractor value \cite{Ooguri:2005vr}, a relation between the attractor
fixed point values and the energy eigenvalues of the 0A matrix model is not unexpected.}.

\section{Black hole entropy and compact Calabi--Yaus}

\label{sec:bhentropy}

We now turn to the relation between matrix models and black hole entropy. In
Ref. \cite{Danielsson:2004ti}, the relation $S_{BH}=-F_{MM}/T_{MM}$ is
derived. $F_{MM}$ is in the canonical ensemble and $T_{MM} = (2 \pi R_{MM})^{-1}$ is the matrix model temperature. 

Matrix models are usually related to topological strings on non-compact
Calabi--Yaus. In the case of a {\em compact} Calabi--Yau the number of independent
three-cycles is $b^{3} = 2(h^{(2,1)}+1)$, and these are naturally divided into symplectic pairs of A- and B-cycles. The complex structure moduli space then has
dimension $h^{(2,1)}$. It can be parameterized by the periods of the holomorphic $(3,0)$-form on $h^{2,1}$ A-cycles, or more invariantly by considering the periods on all $(h^{2,1}+1)$ A-cycles as projective coordinates. For more details, the reader is referred to \cite{Vonk:2005yv} for a review on the special geometry of Calabi--Yaus. 

We would like to think of the {\em non-compact} manifolds as local models for compact Calabi-Yaus, and of the periods of the $n$ A-cycles in these geometries as $n$ true complex structure moduli. This means there has to be at least one extra A-cycle ``at infinity''.  In the formula $S_{BH}=-F_{MM}/T_{MM}$, the left hand side is a function of the black hole charges. Since the number of electromagnetic charges of the four-dimensional black hole equals the total number of three-cycles, it is important to take the extra A-cycle(s) into account when obtaining the dependence of the black hole entropy on the charges in this framework.

In this section we derive an explicit correspondence between the 0A
matrix model at self-dual radius and a four-dimensional half-BPS ${\cal N}=2$ black hole
on a compact internal space with a conifold point. This involves
finding charge configurations that, through the attractor equations
\cite{Ferrara:1995ih, Strominger:1996kf}, fix the moduli to a conifold point
at the horizon. Let us however stress again that this ``decompactification'' only takes place near the black hole horizon, and only for quantities that are sensitive to the complex structure moduli -- the Newton's constant being the most notable exception. We will identify variables, thermodynamical ensembles and double scaling limits on both sides, thus tying up one end left loose in Ref.
\cite{Danielsson:2004ti}. We also verify that the Legendre transforms, taking
us from $\mathcal{F}_{BH}$ to $S_{BH}$ on the black hole side \cite{Ooguri:2004zv}
and from the grand canonical to the canonical ensemble on the matrix model
side, coincide as proposed in Ref. \cite{Danielsson:2004ti}. Finally, we
explore the large classical contributions present on both sides.

For a review on compactification in the black hole context, see Ref.\ \cite{Mohaupt:2000mj}. In \cite{Danielsson:2004ti,Bilal:2005hk}, two different ways to deal with the truly noncompact case by introducing cutoffs were discussed.

\subsection{Charges and decompactification}
Consider for simplicity a Calabi--Yau $\mathcal{M}$ with just one complex
structure modulus. For example, one could think of $\mathcal{M}$ as the mirror quintic, which has been thoroughly studied in \cite{Candelas:1990qd}. Extending the treatment to the general case is straightforward. We
choose a symplectic basis $A^{I}, B_{I}$, $I=0,1$ of $H_{3}(\mathcal{M}, \mathbb{Z})$,
which is such that the period of $A^{1}$ shrinks to zero at the conifold point. Let
$X^{I}$ and $F_{I}$ be the periods of the holomorphic three-form on $A^{I}$ and $B_{I}$.
Each pair of cycles leads to a four-dimensional gauge field, and hence to an electric and a magnetic charge. Our objective is to express the entropy of the black hole as a function of its
electromagnetic charges $q_{I}$ and $p^{I}$.

Recall that the entropy is given by \cite{Ooguri:2004zv}
\begin{equation}
S_{BH}(q_{I},p^{I})=\mathcal{F}_{BH}(\phi^{I},p^{I})-\phi^{J}\frac{\partial
}{\partial\phi^{J}}\mathcal{F}_{BH}(\phi^{I},p^{I}),\label{SBH}%
\end{equation}
where $\phi^{I}$ are the chemical potentials conjugate to $q_{I}$.
$\mathcal{F}_{BH}(\phi^{I},p^{I})$ can be obtained from the topological
partition function $\mathcal{F}_{top}$ on the Calabi--Yau as
\cite{Ooguri:2004zv}
\begin{equation}
\mathcal{F}_{BH}(\phi^{I},p^{I})=2\mbox{Re}\mathcal{F}_{top}(t,g_{top}%
).\label{BH_TS}%
\end{equation}
Here $t=X^{1}/X^{0}$ is a parameter on moduli space, and $g_{top}$ is the
topological string coupling constant. The correspondence holds if
$\mathcal{F}_{top}$ is evaluated at
\begin{equation} \label{eq:BHmoduli}
t=\frac{p^{1}+i\phi^{1}/\pi}{p^{0}+i\phi^{0}/\pi},\mbox{ }g_{top}=\frac{\pm4\pi
i}{p^{0}+i\phi^{0}/\pi}.
\end{equation}
We now want to compute $\mathcal{F}_{BH}$ using matrix model technology. To
this end we use Eq.\ (\ref{FA=Fbos}), and the fact that $\mathcal{F}_{c=1}$ equals the
topological partition function on the conifold \cite{Ghoshal:1995wm}. To be
more specific, in the double scaling limit $t\rightarrow0$, $g_{top}%
\rightarrow0$ at constant $\mu_{top}\equiv t/g_{top}$, the partition function
is given by
\begin{equation}
\mathcal{F}_{c=1}(\mu=\mu_{top})=\operatorname{Re}\mathcal{F}_{top}(i\mu
_{top}).\label{Ftop=FMM}%
\end{equation}
The equality (\ref{Ftop=FMM}) is valid only after appropriately fixing the
gauge on the topological string theory side, and up to large classical terms,
on which we will comment in a moment. Thus, Eqs.\ (\ref{FA=Fbos}), (\ref{BH_TS})
and (\ref{Ftop=FMM}) give $\mathcal{F}_{BH}=\mathcal{F}_{0A}$, upon
identifying variables. This is done by computing the $\sim(q+i\mu)^{2}%
\ln[(q+i\mu)/\beta]$ contribution to the zero order term of $\mathcal{F}%
_{0A}$, and its counterpart $\sim(p^{1}+i\phi^{1}/\pi)^{2}\ln[(p^{1}%
+i\phi^{1}/\pi)/(p^{0}+i\phi^{0}/\pi)]$ in $\mathcal{F}_{BH}$. Doing this
carefully, using e.g. Eq.\ (2.16) of Ref.\ \cite{Ooguri:2004zv}, yields the
identification
\begin{align}
\mu &  \equiv\phi^{1}/\pi,\nonumber\label{variables}\\
q &  \equiv p^{1},\nonumber\\
\beta &  \equiv p^{0}+i\phi^{0}/\pi.
\end{align}
Note that $\beta$ is to be considered as an independent variable. It appears
only in the the genus 0 and 1 contributions, exactly as $p^{0}+i\phi
^{0}/\pi$. Let us stress that $\mu$ is identified with $\phi^{1}$, and not
with $p^{1}$.

With the correspondence (\ref{variables}) we have $\mathcal{F}_{BH} =
\mathcal{F}_{0A}$, and the fact that $\mu\sim\phi^{1}$ shows that the Legendre
transforms on both sides indeed match\footnote{On the black hole side the
transform really contains a $-\phi^{0} \partial\mathcal{F}_{BH}/\partial
\phi^{0}$ term which is not present on the matrix model side. Since $\beta$ only appears in the genus 0 and 1 terms, this
contribution only contains non-universal terms. However, since we will be
interested also in the non-universal terms, we choose the black hole charges
in such a way that $\phi^{0} = 0$ in what follows.}. Thus we have explicitly
verified the conclusion that $S_{BH} = -F_{0A}/T_{SD}$ (Eq.\ (1.2) of Ref.\
\cite{Danielsson:2004ti}), where $F_{0A}$ is the canonical free energy of the
0A matrix model, and $T_{SD} = 1/\pi$ is the self-dual temperature. To be precise, in this ensemble the identification of $q$ and $\beta$ is as in (\ref{variables}), and instead of $\mu$ we now have $N$, which is the number of fermions measured from the top of the $-x^2$ part of the potential. This variable is to be identified with the electric charge $q_{1}$ of the black hole, and has expectation value $\left<{N}\right> =- \frac{1}{\pi}\partial \mathcal{F}_{0A}/\partial\mu$. Having made the connection
(\ref{variables}) we need to identify the double scaling limit on the black
hole side. Indeed, Eq.\ (\ref{Ftop=FMM}) is only valid in that limit, and thus
it is only in this limit that $\mathcal{F}_{0A}$ correctly computes the
entropy. Eq.\ (\ref{eq:BHmoduli}) shows that the appropriate limit is $p^{0}
+ i\phi^{0}/\pi\rightarrow\infty$ while $p^{1} + i\phi^{1}/\pi$ remains
constant\footnote{Note that this limit coincides exactly with the usual matrix
model double scaling limit
\cite{Gross:1990ay}-\cite{Parisi:1989dk}.}. Hence $p^1$ remains constant, and
using the attractor equations, it is straightforward to show that at least two of $p^{0}, q_{0}$ and $q_{1}$
go to infinity. The attractor equations also give the following condition on
these three charges:
\begin{equation}
\label{conifolds}p^{0} \mbox{Im}(F_{0} \bar{F}_{1})+q_{0}\mbox{Im}(F_{1}
\bar{X}^{0})+q_{1}\mbox{Im}(X^{0} \bar{F}_{0})=0,
\end{equation}
where all periods are evaluated at $t=0$. When the charges satisfy these requirements, the Calabi--Yau will be effectively non-compact at the black hole horizon.

\subsection{Classical terms}
Next, let us consider the large classical terms appearing in the black hole
entropy. Computing the genus 0 contribution gives
\begin{equation}
\label{largeterms}\mathcal{F}_{BH} = 2\mbox{Re}\left[  \frac{\pi i}{4}A_{1}
(CX^{0})^{2} + \frac{\pi i}{2} A_{2} (CX^{0})(CX^{1}) + \frac{1}{2}%
(\frac{CX^{1}}{2})^{2}\ln\frac{CX^{1}}{CX^{0}} + \ldots \right]  ,
\end{equation}
where terms that are vanishing or finite and regular in the double scaling limit have been omitted.
Here, $CX^{I} \equiv p^{I} + i\phi^{I}/\pi$, and $A_{i}$ are numerical
constants depending on the Calabi--Yau. Explicitly $A_{1} = (F_{0}%
/X^{0})|_{t=0}$ and $A_{2}=(F_1/X^0)|_{t=0}$. Note that the first two
terms become large in the double scaling limit.

In principle, there are large classical contributions also to the matrix model
free energy. Regularizing the potential of the $c=1$ matrix model $V
\sim-x^{2}$ according to
\begin{equation}
V\left(  x\right)  =-\frac{x^{2}}{\alpha^{\prime}}+A x^{4},
\end{equation}
gives, up to numerical factors, a grand canonical matrix model free energy of
the form
\begin{equation}
\mathcal{F}_{c=1}\left(  \mu,\beta\right)  \sim\frac{1}{A^{2}}\beta^{2}%
-\frac{1}{A}\mu\beta- \mu^{2}\ln\frac{\mu}{\beta} + \ldots
\end{equation}
We see that since we can identify $C X^0 \sim \beta$, $C X^1 \sim i \mu$, the structure of this expression is in complete accordance with (\ref{largeterms}). For the example of the mirror quintic, we also checked that the sign of the leading term is the same in both equations. To precisely match the two undetermined coefficients in (\ref{largeterms}), one would need to consider a potential which is regularized by two coefficients, such as $V \sim -x^2 + A x^4 + B x^6$.

\section{Conclusions and outlook}

In this paper, we have studied the correspondence between matrix models,
topological strings and four-dimensional ${\cal N}=2$ half-BPS black holes. We have in particular studied the relations between these systems for the case of the 0A matrix model at multiples of its self-dual radius. When relating the matrix
models to topological strings, we have argued that it is more natural to match
the matrix model free energy to $2\mbox{Re}\mathcal{F}_{top}$ than
to $\mathcal{F}_{top}$. Consequently, the geometry naturally associated to the
matrix model at self-dual radius is the deformed conifold. This conifold can be viewed as the ``holomorphic square root'' of the manifold that follows from the ground ring equations. At multiples of the self-dual radius we again find such a holomorphic--anti-holomorphic factorization, and the matrix model should be associated with certain non-compact Calabi--Yau manifolds with $n$ three-cycles that can
shrink to zero volume. We found that it is plausible that such local geometries can be embedded in compact Calabi--Yaus.

We noted that the matrix model free energies and the topological partition
functions need not match each other genus by genus. In particular, a mixing of genus
0 and genus 1 terms will occur. This ensures that the coefficients in
front of the genus 1 terms on the topological string side are always $-1/12$
for both the holomorphic and the anti-holomorphic contributions, as required
for the resolution of compact singular spaces by string theory.

Using the recently conjectured correspondence between topological strings and
black holes in type IIB string compactification, we were
able to directly relate the matrix model free energy to the one for the black
hole. An important new point here was that the theory is compactified on a compact Calabi--Yau, that through
the attractor equations develops a singularity only at the black hole horizon. This allowed us to get a matrix model description of the black hole, in spite of the compactness of the Calabi--Yau.

The relation was calculated explicitly for an internal space
with a single conifold point. The variables matched perfectly and we saw that
the Legendre transforms between the canonical and grand canonical ensemble on
the matrix model side, and between $S_{BH}$ and $\mathcal{F}_{BH}$ on the
black hole side were identical. It was also shown that the large classical
terms in ${\cal F}_{BH}$ and ${\cal F}_{MM}$ can be matched in form by regulating the matrix model potential.

There has been much interest in the consequences of the topological string / black hole relation recently, and it seems that many more interesting results in this direction lie ahead. Let us mention some lines of further investigation related to the results of this paper. First of all, it would be extremely interesting to understand the nonperturbative corrections on the different sides of the story better. Many of the models that we have mentioned are perturbatively equivalent, but have nonperturbative differences. Studying these better, as well as their relations to black holes, may give us some intuition about the correct nonperturbative completion of topological string theory, and about the question of how unique such a completion is. A closer nonperturbative study may also lead to relations with the baby universes of \cite{Dijkgraaf:2005bp}. In this respect, the sums over different conifolds we have mentioned are also suggestive.

On a more technical level, the notion of the ``holomorphic square root'' of the ground ring geometry needs to be made more precise. On a case-by-case basis, the correct geometries are not hard to guess, but it seems that by using Kodaira-Spencer theory a more rigorous definition should also be possible.

Another interesting point to work out further would be the actual embedding of the local models into compact Calabi--Yaus. Ultimately, this leads to the intriguing mathematical question of which local Calabi--Yau manifolds allow an embedding into compact Calabi--Yaus. Already in the two-moduli case this seems to be a very nontrivial issue.

Finally, we repeat an open question that was mentioned in \cite{Danielsson:2004ti}: could we completely skip the topological string step and directly relate the matrix models to the black holes? Since the black holes have an $AdS_2 \times S^2$ near-horizon region, one would expect the supergravity theory to be equivalent to a $CFT_1$ on the boundary of $AdS_2$. It seems natural to relate the two factors of the matrix model partition function to the two boundaries of this space. It would be interesting to make such a holographic description precise.

\section*{Acknowledgments}

UD is a Royal Swedish Academy of Sciences Research Fellow supported by a grant
from the Knut and Alice Wallenberg Foundation. The work was supported by the
Swedish Research Council (VR) and the Royal Swedish Academy of Science.


\begin{thebibliography}{99}                                                                                               %

\bibitem {Ooguri:2004zv}H.~Ooguri, A.~Strominger and C.~Vafa, ``Black hole
attractors and the topological string,'' Phys.\ Rev.\ D \textbf{70}, 106007
(2004) [arXiv:hep-th/0405146].

\bibitem {Ghoshal:1995wm}D.~Ghoshal and C.~Vafa, ``C = 1 string as the
topological theory of the conifold,'' Nucl.\ Phys.\ B \textbf{453}, 121 (1995)
[arXiv:hep-th/9506122].

\bibitem{Witten:1991zd}
E.~Witten, ``Ground ring of two-dimensional string theory,''
Nucl.\ Phys.\ B {\bf 373}, 187 (1992) [arXiv:hep-th/9108004].

\bibitem {Aganagic:2003qj}M.~Aganagic, R.~Dijkgraaf, A.~Klemm, M.~Marino and
C.~Vafa, ``Topological strings and integrable hierarchies,''
arXiv:hep-th/0312085.

\bibitem {Ita:2004yn}H.~Ita, H.~Nieder, Y.~Oz and T.~Sakai, ``Topological
B-model, matrix models, c-hat = 1 strings and quiver gauge theories,'' JHEP
\textbf{0405}, 058 (2004) [arXiv:hep-th/0403256].

\bibitem {Hyun:2005fq}S.~Hyun, K.~Oh, J.~D.~Park and S.~H.~Yi, ``Topological
B-model and $\hat{c} = 1$ string theory,'' arXiv:hep-th/0502075.

\bibitem {Witten:1993ed}E.~Witten, ``Quantum background independence in string
theory,'' arXiv:hep-th/9306122.

\bibitem {Dijkgraaf:2002ac}R.~Dijkgraaf, E.~Verlinde and M.~Vonk, ``On the
partition sum of the NS five-brane,'' arXiv:hep-th/0205281.

\bibitem{Vonk:2003th}
M.~Vonk, ``Dual perspectives on extended objects in string theory,'' PhD-thesis,
University of Amsterdam (2003).

\bibitem {Danielsson:2004ti}U.~H.~Danielsson, M.~E.~Olsson and M.~Vonk,
``Matrix models, 4D black holes and topological strings on non-compact
Calabi-Yau manifolds,'' JHEP \textbf{0411}, 007 (2004)
[arXiv:hep-th/0410141].

\bibitem {Takayanagi:2003sm}T.~Takayanagi and N.~Toumbas, ``A matrix model
dual of type 0B string theory in two dimensions,'' JHEP \textbf{0307}, 064
(2003) [arXiv:hep-th/0307083].

\bibitem {Douglas:2003up}M.~R.~Douglas, I.~R.~Klebanov, D.~Kutasov,
J.~Maldacena, E.~Martinec and N.~Seiberg, ``A new hat for the c = 1 matrix
model,'' arXiv:hep-th/0307195.

\bibitem {Vafa:1995ta}C.~Vafa, ``A Stringy test of the fate of the conifold,''
Nucl.\ Phys.\ B \textbf{447}, 252 (1995) [arXiv:hep-th/9505023].

\bibitem {Ferrara:1995ih}S.~Ferrara, R.~Kallosh and A.~Strominger,
\textquotedblleft N=2 extremal black holes,\textquotedblright\ Phys.\ Rev.\ D
\textbf{52}, 5412 (1995) [arXiv:hep-th/9508072].

\bibitem {Strominger:1996kf}A.~Strominger, ``Macroscopic Entropy of $N=2$
Extremal Black Holes,'' Phys.\ Lett.\ B \textbf{383}, 39 (1996)
[arXiv:hep-th/9602111].

\bibitem {Maldacena:2005he}J.~Maldacena and N.~Seiberg, ``Flux-vacua in Two
Dimensional String Theory,'' arXiv:hep-th/0506141.

\bibitem {Jevicki:1993zg}A.~Jevicki and T.~Yoneya, ``A Deformed matrix model
and the black hole background in two-dimensional string theory,''
Nucl.\ Phys.\ B \textbf{411}, 64 (1994) [arXiv:hep-th/9305109].

\bibitem {Danielsson:1993wq}U.~H.~Danielsson, ``A Matrix model black hole,''
Nucl.\ Phys.\ B \textbf{410} (1993) 395 [arXiv:hep-th/9306063].

\bibitem {Demeterfi:1993sj}K.~Demeterfi and J.~P.~Rodrigues, ``States and
quantum effects in the collective field theory of a deformed matrix model,''
Nucl.\ Phys.\ B \textbf{415}, 3 (1994) [arXiv:hep-th/9306141].

\bibitem {Demeterfi:1993cm}K.~Demeterfi, I.~R.~Klebanov and J.~P.~Rodrigues,
``The Exact S matrix of the deformed c = 1 matrix model,''
Phys.\ Rev.\ Lett.\ \textbf{71}, 3409 (1993) [arXiv:hep-th/9308036].

\bibitem {Danielsson:1994ac}U.~H.~Danielsson, \textquotedblleft
Two-dimensional string theory, topological field theories and the deformed
matrix model,\textquotedblright\ Nucl.\ Phys.\ B \textbf{425} (1994) 261. [arXiv:hep-th/9401135].

\bibitem {Danielsson:1994sk}U.~H.~Danielsson, \textquotedblleft The Scattering
of strings in a black hole background,\textquotedblright\ Phys.\ Lett.\ B
\textbf{338} (1994) 158 [arXiv:hep-th/9405052].

\bibitem{Danielsson:2003yi}
U.~H.~Danielsson, ``A matrix model black hole: Act II,''
JHEP {\bf 0402}, 067 (2004) [arXiv:hep-th/0312203].

\bibitem {Gukov:2003yp}S.~Gukov, T.~Takayanagi and N.~Toumbas, ``Flux
backgrounds in 2D string theory,'' JHEP \textbf{0403} (2004) 017
[arXiv:hep-th/0312208].

\bibitem{Danielsson:2004xf}
U.~H.~Danielsson, J.~P.~Gregory, M.~E.~Olsson, P.~Rajan and M.~Vonk,
``Type 0A 2D black hole thermodynamics and the deformed matrix model,''
JHEP {\bf 0404}, 065 (2004) [arXiv:hep-th/0402192].

\bibitem {Strominger:1995cz}A.~Strominger, ``Massless black holes and
conifolds in string theory,'' Nucl.\ Phys.\ B \textbf{451}, 96 (1995)
[arXiv:hep-th/9504090].

\bibitem {Danielsson:1993dh}U.~H.~Danielsson, ``The deformed matrix model at
finite radius and a new duality symmetry,'' Phys.\ Lett.\ B \textbf{325}, 33
(1994) [arXiv:hep-th/9309157].

\bibitem {Kapustin:2003hi}A.~Kapustin, ``Noncritical superstrings in a
Ramond-Ramond background,'' JHEP \textbf{0406}, 024 (2004)
[arXiv:hep-th/0308119].

\bibitem {Gopakumar:1998vy}R.~Gopakumar and C.~Vafa, ``Topological gravity as
large N topological gauge theory,'' Adv.\ Theor.\ Math.\ Phys.\ \textbf{2}
(1998) 413 [arXiv:hep-th/9802016].

\bibitem {Bershadsky:1993cx}M.~Bershadsky, S.~Cecotti, H.~Ooguri and C.~Vafa,
``Kodaira-Spencer theory of gravity and exact results for quantum string
amplitudes,'' Commun.\ Math.\ Phys.\ \textbf{165}, 311 (1994)
[arXiv:hep-th/9309140].

\bibitem {Ooguri:2005vr}H.~Ooguri, C.~Vafa and E.~Verlinde,
``Hartle-Hawking wave-function for flux compactifications,''
arXiv:hep-th/0502211.

\bibitem {Vonk:2005yv}M.~Vonk, ``A mini-course on topological strings,''
arXiv:hep-th/0504147.

\bibitem {Mohaupt:2000mj}T.~Mohaupt, ``Black hole entropy, special geometry
and strings,'' Fortsch.\ Phys.\ \textbf{49}, 3 (2001) [arXiv:hep-th/0007195].

\bibitem{Bilal:2005hk}
A.~Bilal and S.~Metzger, ``Special geometry of local Calabi-Yau manifolds and superpotentials from
holomorphic matrix models,'' arXiv:hep-th/0503173.

\bibitem {Candelas:1990qd}P.~Candelas, X.~C.~De la Ossa, P.~S.~Green and
L.~Parkes, ``An Exactly Soluble Superconformal Theory From A Mirror Pair Of
Calabi-Yau Manifolds,'' Phys.\ Lett.\ B \textbf{258}, 118 (1991).

\bibitem{Gross:1990ay}
D.~J.~Gross and N.~Miljkovic, ``A Nonperturbative Solution Of D = 1 String Theory,''
Phys.\ Lett.\ B {\bf 238}, 217 (1990).

\bibitem{Brezin:1989ss}
E.~Brezin, V.~A.~Kazakov and A.~B.~Zamolodchikov,
``Scaling Violation In A Field Theory Of Closed Strings In One Physical
Dimension,'' Nucl.\ Phys.\ B {\bf 338}, 673 (1990).

\bibitem{Ginsparg:1990as}
P.~H.~Ginsparg and J.~Zinn-Justin, ``2-D Gravity + 1-D Matter,''
Phys.\ Lett.\ B {\bf 240}, 333 (1990).

\bibitem{Parisi:1989dk}
G.~Parisi, ``On The One-Dimensional Discretized String,''
Phys.\ Lett.\ B {\bf 238}, 209 (1990).

\bibitem{Dijkgraaf:2005bp}
R.~Dijkgraaf, R.~Gopakumar, H.~Ooguri and C.~Vafa, ``Baby universes in string theory,''
arXiv:hep-th/0504221.

\end{thebibliography}
\end{document}